\documentstyle[11pt,newpasp,twoside,epsf]{article}
\def\mfig #1#2#3#4{\par
\epsfxsize=#1 cm
\moveright #2cm
\vbox{\epsfbox{#3}}
{\noindent Figure~#4 }\vskip .3cm }

\def\myfig #1#2#3#4{
\vskip #1 cm
\epsfxsize=#1 cm
\moveright #2cm
\vbox{\epsfbox{#3}}
{\noindent Figure~#4 }\vskip .3cm 
}
\def\myfig #1#2#3#4{
\begin{figure}
\plotone{#3}
\caption{#4}
\end{figure}}

\def\lg{{\rm log}}
\def\bh{black hole~}
\def\bhs{black holes~}

\def\el {emission-line}

\def\ers{{\rm erg/sec}}
\def\ms{M_{\odot}}
\def\et{et al.\ }

\def\rev{reverberation~}
\def\vFWHM{\ifmmode v_{\mbox{\tiny FWHM}} \else
            $v_{\mbox{\tiny FWHM}}$\fi}
\def\kms{\ifmmode {\rm km\ s}^{-1} \else km s$^{-1}$\fi}
\def\ers{\ifmmode {\rm erg\ s}^{-1} \else erg s$^{-1}$\fi}

\def\apj{{\it Astrophysical Journal~}}
\def\apjl{{\it Astrophysical Journal (Letters)}}

\def\aa{{\it Astronomy and Astrophysics}}

\def\edcomment#1{\iffalse\marginpar{\raggedright\sl#1\/}\else\relax\fi}
\reversemarginpar

\begin{document}
\title{The Black Hole-Host Galaxy Relation in Active Galactic Nuclei}
\author{Amri Wandel}
\affil{Racah Inst. of Physics, The Hebrew University, Jerusalem 91405,Israel}

\begin{abstract}
The masses of the central black holes of Active Galactic Nuclei estimated by the reverberation virial 
technique seem to be significantly smaller than predicted from the black hole mass-bulge 
luminosity relation for massive black holes detected by kinematic methods in less active and normal galaxies.
We reexamine the relation for updated data and suggest a few reasons that may cause this discrepancy.
\end{abstract}

\section{Introduction}

Compact dark masses, probably massive \bhs (MBHs), have been detected in the cores of 
many normal galaxies  (Kormendy and Richstone 1995).
The \bh mass seems to correlate with the host galactic
bulge, although with a significant scatter,
with the MBH being
 about 0.006 of the mass of the spheroidal bulge 
(Magorrian \et 1998). 
Recent research using higher quality HST data and a 3-integral model give 
lower \bh masses and an average \bh to bulge mass ratio of 0.002
(Ho 1999).

In addition to the MBHs detected by techniques of stellar and gas kinematics,
recently the masses of  MBHs in AGNs has been estimated  by reverberation mapping 
of the broad emission-line region and the virial assumption.
High quality reverberation data and virial masses are presently available 
for 20 Seyfert 1 nuclei (Wandel, Peterson and Malkan 1999, hereafter WPM) 
and 17 PG quasars (Kaspi \et 2000).
It is interesting to find whether AGNs follow a similar \bh -bulge relation  
as normal galaxies. This could test of the generality of this relation 
as well as reveal new facts bout the nature of AGNs, in particular about the
relation  between the host galaxy and the central source. 

Evidence that the \bh is related with the virial mass on an intermediate
scale, the narrow line region, comes from the correlation between the \bh mass
and the velocity dispersion of the narrow emission-line gas, as measured by the 
OIII lines in Seyfert 1 nuclei (Wandel and Mushotzky 1986; Nelson 2000).

For Seyfert 1 galaxies the measurement of the \bh mass and the host bulge
is easier than in more luminous AGNs: Because of their lower nuclear brightness,
the bulges of Seyfert galaxies can be observed 
more easily and closer to the center than in quasars. 
In addition, the BLR in the less luminous Seyferts is smaller and hence reverberation 
mapping requires a shorter time sequence.

Using the WPM sample Wandel (1999) found that 
the ratio  of \bh mass to bulge magnitude for  Seyfert 1s is significantly lower (by a 
factor of 20, on average) than the ratio found by  Magorrian  \et (1998) for 
MBHs in normal galaxies. 
 Comparing the reverberation masses with the more careful analyses of stellar 
dynamics, Ho (1999) found lower \bh masses in normal galaxies, reducing the 
discrepancy with the \rev -mapped Seyferts, but a factor of $\sim 5$ remained. 

Recently it has been found that the correlation of the \bh mass with the stellar 
velocity dispersion (rather than the bulge luminosity) is significantly tighter 
than with the bulge luminosity (Ferarrese and Merrit 2000, Gebhardt \et 2000a).
Furthermore, in the \bh mass - stellar velocity dispersion plane, the discrepancy 
between normal galaxies and Seyferts seems to disappear (Gebhardt \et 2000b). 
In this work we suggest some possible explanations to these observations.

\section{BLR reverberation as a probe of \bh masses}

The best resolution of HST in relatively nearby galxies translates into
a distance of a few tens of parsecs from the MBH. Reverberation mapping 
of the broad emission line region in
AGNs gives a much closer view - a few light days from the center. 
Assuming the line-emitting matter is gravitationally bound, 
having a Keplerian velocity dispersion , 
it is possible to estimate the virial mass:
$M\approx G^{-1}rv^2 .$
 This expression may be approximately valid also in the case
the line emitting gas is not bound,
such as radiation-driven motions and disk-wind models (e.g. Murray \et 1998).
The main problem in estimating the virial mass from the \el~ data is to obtain
a reliable estimate of the size of the BLR, and to correctly identify the
line width with the velocity dispersion in the gas. 
WPM used the continuum/emission-line cross-correlation function to measure the
responsivity-weighted radius $c\tau$ of the BLR (Koratkar \& Gaskell 1991), 
and the variable (rms) component of the spectrum to measure the velocity dispersion 
in the same
part of the gas which is used to calculate the BLR size, automatically 
excluding constant features such as narrow emission lines and 
Galactic absorption.
The case for a central MBH dominating the kinematics in the broad emission-line region
is supported by the Keplerian velocity profile ($v \propto r^{-1/2}$) detected in
NGC 5548 
(Peterson \& Wandel 1999). 
The virial masses derived from different emission lines at several different epochs are 
all consistent with a single value ($(6.3\pm 2)\times 10^7\ms$)
which demonstrates the case for a
Keplerian velocity dispersion in the line-width/time-delay data.
A similar result holds for three other AGNs (Peterson \& Wandel 2000).

\section {The Black-Hole - Bulge Relation }
\subsection{Seyfert 1 galaxies}

The \bh -bulge relation has been derived for the Seyfert 1 galaxies in 
 the WPM sample with the virial mass derived from the H$\beta$ line by 
the reverberation-rms method. The bulge magnitudes were taken
from the compilation of Whittle \et (1992), who calculate
the bulge magnitude from the total blue magnitude, using 
the empirical formula of Simien \& deVaucolours (1986), relating the galaxy
type to the bulge/total fraction. The bulge magnitudes are corrected for the nonstellar emission using the correlation between H$\beta$ and the 
nonstellar continuum luminosity (Shuder 1981). 

Fig. 1 shows the \bh~ mass as a function of the bulge luminosity for
all the objects in Wandel (1999) plus NGC3516 and NGC 4593.
The best fit is
$\lg M_{bh} = 1.1 \lg L_{bulge} - 3.9$
where the \bh mass and bulge luminosity are in solar units.
This is by a factor of $\sim 5-7$ lower compared to the 
equivalent relation for MBHs in normal galaxies (fig 2 of Gebhardt \et 2000a,
represented by the dashed line in fig. 1). This difference is significant even 
when the uncertainties  
in the bulge luminosity and \bh mass  are considered.

Translating the bulge luminosity to mass (using the mass-to-light relation for normal
galaxies, 
${M/\ms \over L/L_\odot}~\approx 5 (L/10^{10}L_\odot )^{0.15}$, Faber \et 1997) 
the Seyfert sample average is
$<M_{BH}>= 0.0003 <M_{bulge}>$, significantly lower than 0.002, the value
 for MBHs in normal galaxies (Ho 1999).

\begin{figure}
\plotfiddle{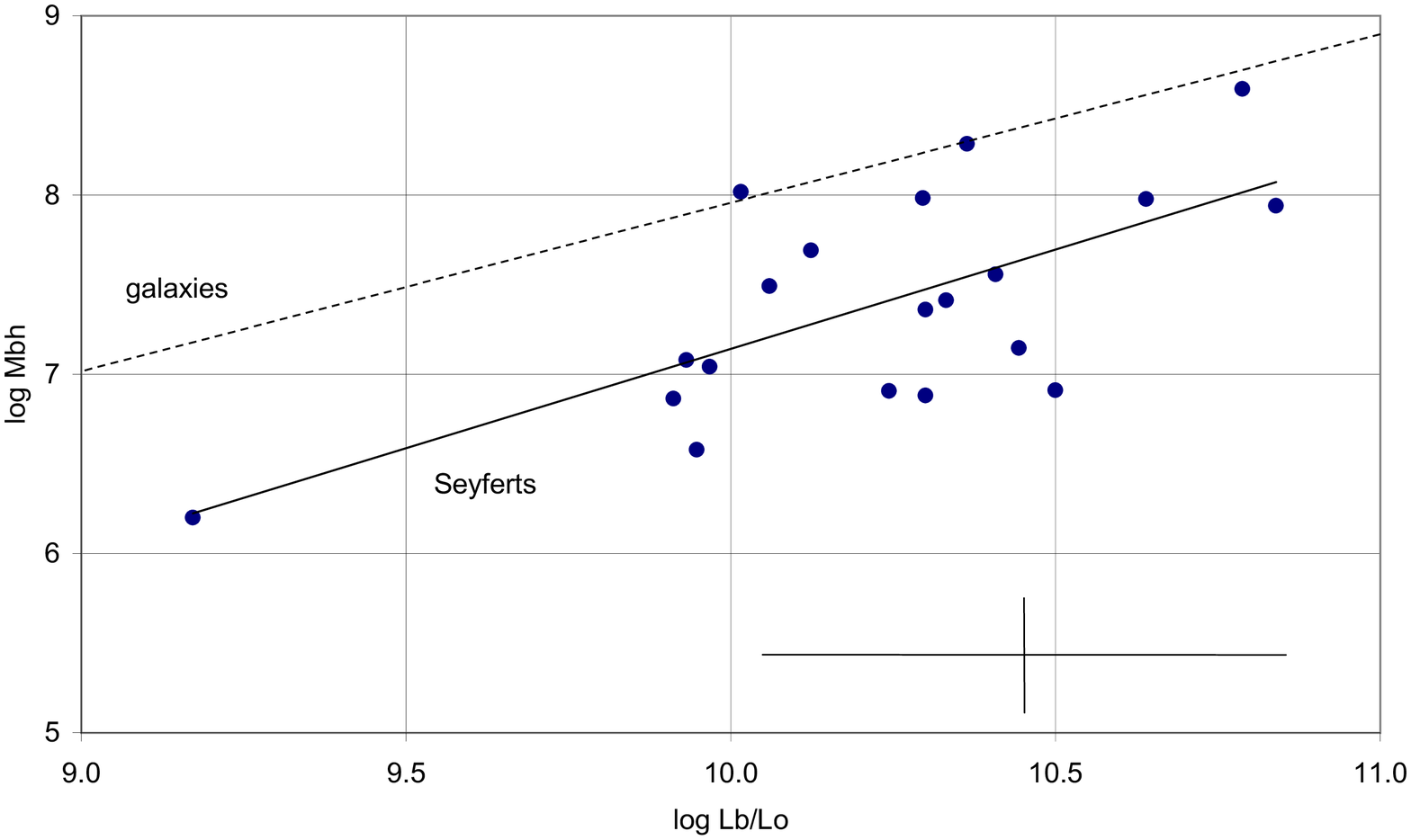}{450pt}{0}{55}{95}{-240}{-90}
\caption
{The virial \bh ~mass calculated by the reverberation BLR
method (from Wandel, Peterson \& Malkan 1999) vs. the bulge luminosity. 
Typical uncertainties are represented by the cross in the lower right part. 
The  diagonal lines are the best fit for normal galaxies
(dashed, Gebhardt \et 2000a) and Seyfert 1 galaxiess 
(solid, Wandel 1999 and this work).
}
\end{figure}

\subsection {Quasars}

Laor (1998) studied the \bh -host bulge relation for a sample
of 15 bright quasars.
The bulge luminosity is estimated from Bahcall \et (1997)  
is rather uncertain as the host galaxy is
dominated by the much brighter nonstellar source.
Also the \bh~ mass is estimated using the empirical radius-luminosity relation 
$r_{BLR}=15 L_{44}^{1/2}~ ~{\rm light-days}$
where $L_{44}=L(0.1-1\micron )$ in units of $10^{44}\ers$.
As this relation has been derived for less than a dozen low- and medium 
luminosity objects (mainly Seyferts) with measured reverberation sizes,
it is not obvious that it may be extrapolated to more luminous quasars.
One could test the empirical approximation by comparing the masses to the 
reverberation masses measured by Kaspi \et (2000), who find a different $r-L$ 
relation. Unfortunately there are only four common quasars in the two samples. 
For three of them the reverberation masses are lower (by factors of 1.5-3) than 
the masses estimated by Laor.

\subsection {Comparing Normal Galaxies, Seyferts and Quasars}

Fig. 2 shows the three groups in the plane of \bh~mass vs. 
bulge luminosity.
The best fits to the 
data in the three groups are
($M_8=M_{BH}/10^8\ms$ and $L_{10}=L_{bulge}/10^{10}L_\odot$):
\begin{enumerate}
\item
Galaxies (Gebhardt \et 2000a) - 
$\log M_8= 1.14 \log L_{10} -0.1$, $ R^2  = 0.81$
\item
Quasars (Laor 1998) - 
$\log M_8= 1.1 \log L_{10} + 0.2$, $ R^2  = 0.52$
\item
Seyferts (WPM) -
$\log M_8= 1.1 \log L_{10}-0.8$, $R^2 = 0.45$
\end{enumerate}
We see that the three slopes are quite similar (and, within the uncertainty of 
the fit, consistent with unity);
Quasars are consistent with normal galaxies (if the masses are decreased
by 0.3 dex, as may be concluded by the comparison with the reverberation values,
the two fits almost overlap). Seyfert galaxies are by a factor of 5 lower.

\begin{figure}
\plotfiddle{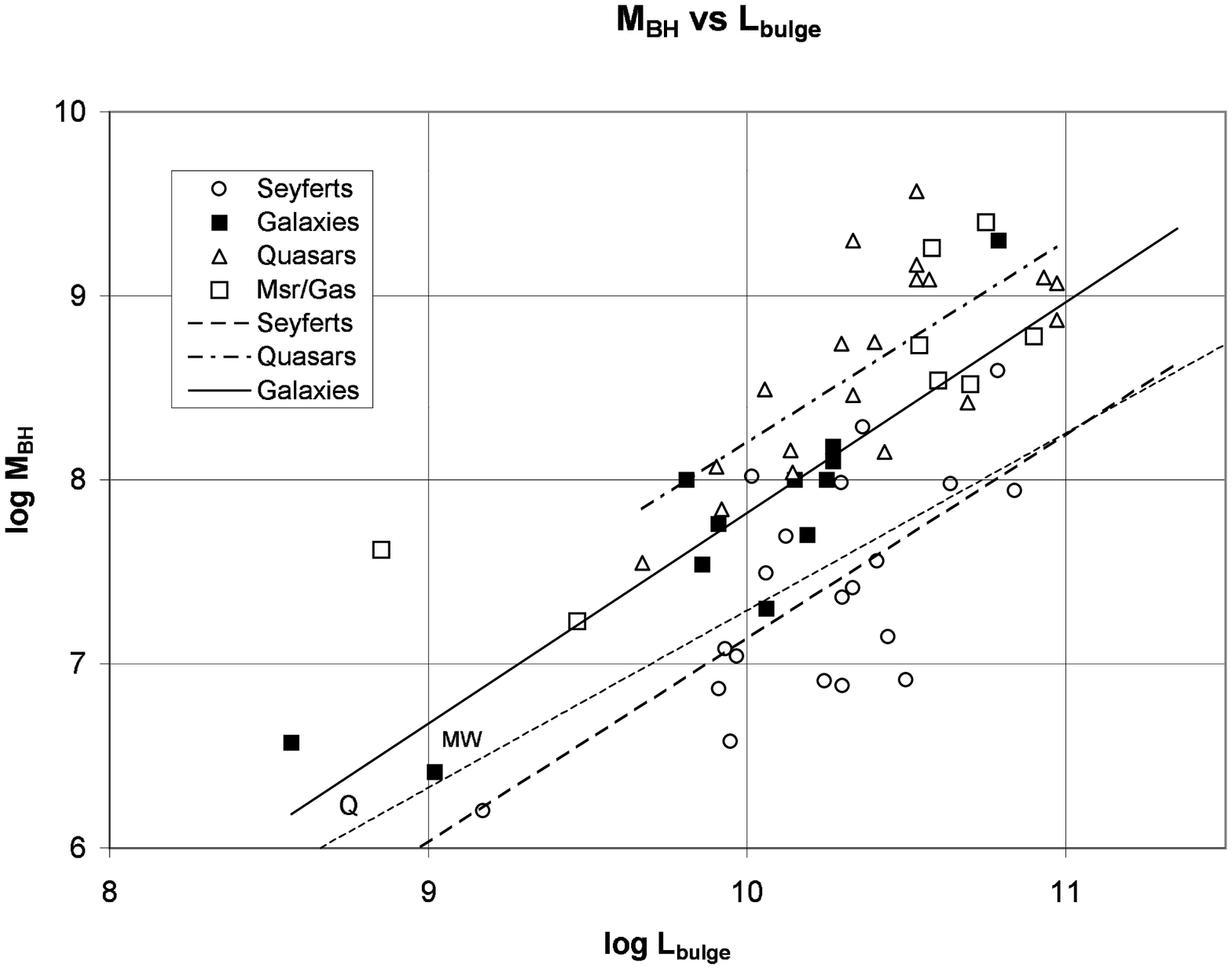}{400pt}{0}{65}{95}{-240}{-80}
\caption{
Mass estimates of MBHs  plotted against the
luminosity of the host galaxy (or the bulge for spiral galaxies).
Squares: MBH in normal galaxies (solid- stellar dynamics, 
open- maser and gas dynamics), 
triangles - 
PG quasars, circles - Seyfert 1 galaxies.
MW denotes our Galaxy.
Also given are the best linear fits for each class (see text).
The dashed line (Q) denotes an estimate of dead \bhs from integrated AGN light.
}
\end{figure}

As a group Seyfert 1 galaxies have a significantly lower \bh to bulge luminosity ratios
than normal galaxies and quasars.
This lower value agrees with the remnant \bh~ density derived from
integrating the emission from quasars (the dashed line denoted Q in fig 2). 
Compared to the density of starlight in galaxies ( $\rho_{gl}$) it gives 
(Chokshi and Turner 1992)
  $\rho_{BH}/\rho_{gl}= 2\times 10^{-3} (0.1/\epsilon )(\ms /L_\odot )$
where  $\epsilon$ is the efficiency.

\section{The \bh mass -velocity dispersion relation for AGN}
The \rev \bh masses of Seyfert galaxies seem to be significantly smaller, when 
compared to the host  bulge luminosity, than MBHs in normal galaxies.
However, compared to the stellar velocity dispersion the difference is much 
smaller or nonexisting (Gebhardt \et 2000b; Merritt \& Ferrarese 2000).
The agreement between Seyferts and normal galaxies in the \bh mass-velocity 
dispersion relation suggests that the \rev masses are essentially correct. 
Nevertheles, this relation remains
 to be confirmed for a larger number of AGNs
(at present  only seven of the Seyfert galaxies with \rev mapping have
reliable determinations of stellar velocity dispersion. In particular, the Seyfert
galaxy NGC5548 has not been included in the sample of Gebhardt \et (2000b) 
because it had a poor
velocity dispersion (Nelson and Whittle 1995). If it were included, it would violate
the \bh -velocity dispersion relation by a large factor, requiring a {\it 3 times larger}
velocity dispersion than measured, in order to fit. 

Another complication could be that the \bh is influencing the stellar velocity 
dispersion measured
for the more massive \bhs . 
The influence of the MBH on the stellar velocity dispersion
can be estimated by comparing the expected
velocity field $(v\propto (M_{\rm BH}/r)^{1/2}$) due to the MBH
and the velocity dispersion of the bulge of the host galaxy.
The \bh enhances the velocity dispersion in an observed region 
if its mass is comparable to the stellar mass
within the radius corresponding to the projected angular size.
This can be measured by
the dimensionless quantity
$$m = GM_{\rm BH}/\sigma^2 \theta D,$$
where $M_{\rm BH}$ is the mass of the MBH,
$\sigma$ is the ``background'' stellar velocity
dispersion in the bulge of the host galaxy,
$\theta$ is the angular size (e.g. the width of the slit) and
$D$ is the distance to the galaxy.
In other words, $m$ is roughly the ratio of the mass of the
MBH to the stellar bulge
mass inside the radius corresponding to the angular size being sampled.
The nuclear velocity dispersion in normal galaxies is typically sampled 
with a slit of order 1-2\arcsec .
Since the measured velocity dispersion is line-of sight 
brightness-weighted, and since the brightness-density increases steeply at
the center, the effective value of $\theta$ is probably even smaller.
Quantitatively, the influence parameter can be written as
$m= 0.9 (M/10^8 M_\odot )
(\sigma/100 km/s)^{-2}(D/10 Mpc)^{-1}$,
where $\theta$ has been assumed to be $1\arcsec$.
Typically $\sigma =150-300\kms$, $M_{bh}\sim 10^7-10^8 M_\odot$
and $D\sim 10-20 Mpc$, hence $m$ for the galaxies with measured velocity dispersion and
\bh mass (the samples of Gebhardt \et 2000a and Ferrarese \& Merrit 2000) 
is of order 0.01-1, with the more massive \bhs
having larger values. If $m$ is larger than  0.2-0.3, the observed nuclear velocity dispersion 
is noticably enhanced. This effect is weaker in AGN, because the contribution from the 
inner part must be blocked out or subtracted, hence the effective $\theta$ is larger
than in normal galaxies, the velocity dispersion is smaller relatively to the 
bulge luminosity, hence ``shifting'' AGNs to lower velocity dispersions (relatively to 
e.g. the bulge luminosity) making them appear to agree better with the $M_{bh}-\sigma$ 
relation of ``normal galaxies'' than with their $M_{bh}-L_{bulge}$ relation. 

\section{Is the difference between Seyferts and normal galaxies apparent or real?}

Figures 1 and 2 show that in spite of the scatter there is a significant difference in 
the \bh mass- bulge luminosity (or mass) relationship of Seyferts and ordinary galaxies,
the former showing systematically lower $M_{bh}/L_{bulge}$
values. This may be an apparent effect, 
due to one of several reasons: (i) \rev masses in Seyferts are 
systematically too low, (ii) the measured bulge luminosity in Syeferts is 
systematically too large, (iii) \bh masses measured by kinematical methods
 are systematically too large. These possibilities are considered
in the following sections. 
There could be a forth cause, that the difference between the two groups is real:
either Seyfert galaxies have systematically larger (or brighter) bulges, or
they have systematically smaller \bhs. The latter case is discussed in the last section.
There are a number of observational errors or uncertainties 
that may cause an apparent difference in the
\bh - bulge luminosity relation between AGN and normal galaxies: 
\subsection{Errors in Seyferts' bulge luminosities}

The discrepancy between Seyferts and normal galaxies' MBH-bulge relation
could be accounted for if the bulge luminosities of the Seyfert galaxies 
are systematically over estimated.
The recent result that  Seyfert nuclei are in good agreement with normal galaxies'
MBHs when the velocity dispersion, rather than the bulge luminosity is being
considered (Gebhardt 2000b, Merritt and Ferrarese 2000) suggests that the problem 
may be in the bulge magnitude determination for Seyferts.
However,  this explanation requires that the bulge magnitude of Seyferts be
{\it systematically} over-estimated by a factor of 3-10 for most Seyfert galaxies
in the sample. As the uncertainty in the determination of the 
bulge magnitude and in the deduction of the non-stellar component 
is significantly less (Whittle 1992), such a large and systematic
over-estimation is not likely. 

\subsection{Systematic errors in the \bh~masses}
It has been shown that the \bh masses in Magorrian \et (1998), especially at the high-mass end of the distribution are overestimated. While the difference between the BH-bulge luminosity relations of Magorrian \et and the Seyfert reverberation sample was a factor of 20 (Wandel 1999), with the corrected sample
it decreases to 5 (fig 2). 
Considering the careful determination of the corrected \bh mass in normal 
galaxies, a further significant decrease in unlikely.

\subsection{Uncertainties and non-virial BLR dynamics}
Is it possible that the Seyfert \bh -\rev masses are systematically underestimated by 
the remaining factor? 
This is unlikely as the \rev masses seem to be more or less correct, in light of  the 
good agreement in the case of the \bh - velocity dispersion relation.
The uncertainty in the \rev virial method is not well known. While for individual objects 
a factor of 2 is probably representative, the sample average error is probably much smaller.
Note  that if the virial assumption is incorrect and the BLR is unbound, 
the gas velocity is actually larger
than Keplerian, the \bh mass would have been {\it overestimated}, which increases the 
discrepancy between the \bh -bulge relation in Seyfert galaxies and
normal galaxies. 

\subsection{Bias introduced by the stellar kinematics method}

Is it possible that the Seyfert galaxies in the WPM sample represent a larger population 
of galaxies with low \bh -to bulge luminosity ratios, which is under-represented in
the presently available MBH sample? This could be the case for MBHs detected using stellar 
dynamics, because in this method cannot detect small \bhs , and the detection-limit 
increases with distance. In a resolution-limited method, this would infer a detection 
lower limit which increases with luminosity.
This hypothesis is supported by the distribution of \bh masses in fig. 2:
there are only few MBHs with masses under 
~$ 10^8\ms$ detected by stellar dynamical methods. These are the Milky Way M31 and its 
satellite M32, and NGC 3377. These galaxies
 do have low \bh -to bulge luminosity ratios, comparable to the Seyfert average relation.
In angular-resolution limited methods, the MBH detection limit 
is correlated with bulge luminosity: for more luminous
bulges the detection limit is higher, because the stellar velocity
dispersion is higher (the Faber-Jackson relation). In order to detect the 
dynamic effect of a MBH it is necessary to observe closer to the center,
while the most luminous galaxies tend to be at  larger distances, so for
a given angular resolution, the MBH detection limit is higher.
This may imply that the sample is biased towards larger
MBHs, as present stellar-dynamical methods are ineffective
for detecting MBHs below $\sim 3~10^7 \ms$ (except in the nearest galaxies).
The BLR method is not subject to this constraint,
hence Seyfert galaxies may be revealing the relatively
low-mass MBH population.

\section{Orientation and BLR geometry}

If the broad emission-line region (BLR) has a flattened geometry, the velocity inferred from the 
observed line width may smaller than the 3D velocity dispersion, depending on the 
inclination and the amount of flattening. For example, for a flat BLR configuration 
viewed at an  inclination angle $i$ (where $i=0$ would be pole-on), the line-of-sight 
velocity is smaller than the 3-D velocity by a factor of sin($i$). 
The inferred mass would be smaller than the actual mass by a factor of $\sin^2(i)$. For a 
random distribution of inclination angles, the average inferred mass would be decreased 
by this factor weighted by the distribution, 
$<\sin^2(i)>=4\pi\int_0^{\pi/2} \sin^2(i)\cos(i){\rm d}i/4\pi=1/3$.
Actually, the distribution is probably 
not random. According to the unified scheme, Seyfert 1 nuclei are viewed more pole-on, 
within an openning angle of $\la 60\deg$  so that the weighted
 average mass-reduction factor is 
$<\sin^2(i)>=4\pi\int_0^{\pi/6} \sin^2(i)\cos(i){\rm d}i/2\pi=1/12$.
If the BLR distribution 
is not flat but with an angular distribution of orbits spanning an angle $\delta$, the effect 
would be less - between sin($i$) and
sin$(\delta+i)$, depending on the distribution of orbits,
$<\sin^2(i+\delta)>~\ga 0.1$. 
This can easily explain the factor $\sim$5 discrepancy between the \bh mass - bulge 
luminosity relation in normal galaxies and reverberation AGNs, even for mildly flattened 
BLR geometry. 
The orientation effect also explains the significantly larger scatter of the Seyferts, 
compared with normal galaxies - due to the distribution of inclination angles.

\section{Quasar-Seyfert Discrepancy: Evolution?}

The arguments above could explain the discrepancy between MBHs in normal galxies and 
Seyferts. Since they can all be applied also to quasars, one could expect quasars 
to show a similar discrepancy compared to MBHs in normal galaxies.
However, this is not the case, as quasars appear to have a similar \bh -to bulge 
luminosity relation as normal galaxies. Hence, there
is a discrepancy between quasars and Seyferts, which cannot be explained by the
arguments in the previous sections. 

One possibility to ``reconcile'' the \bh -bulge relation of the quasar sample 
with the Seyferts' one would be that  the bulge luminosity in Laor's work 
is systematically 
{\it under}estimated by a factor of $\sim$10. Laor (1998) admits the
difficulty in estimating the bulge component, but estimates the uncertainty 
to be modest and not systematic. As noted, 
three of the four PG quasars in Laor's sample which have  
reverberation data have masses  lower (by factors of 1.5-3) than 
the masses estimated by Laor. This factor is however not enough to explain the
quasar-Seyfert difference.

Another explanation is that the difference between the \bh-bulge relation in
Seyferts and quasars is intrinsic.  Wandel (1999) suggested that during their
very luminous phase AGN approach the Eddington luminosity and the central \bh 
accretes within a relatively short timescale a significant fraction of the available 
matter, reaching higher masses (compared to the bulge mass). While Seyferts have 
presumably not yet reached that very luminous phase, MBHs in normal galaxies
may be the remnants of a luminous AGN phase. This scenario could explain 
both: the discrepancy between Seyferts and quasars, and the matching BH-bulge relations
of quasars and MBHs in normal galaxies.
This has been followed by a simple model 
 calculation  by Wang, Biermann and Wandel (2000) who 
examined the relative contribution of mergers to the bulge and MBH masses. For reasonable
parameters they find that the bulge to \bh mass ratio of $\sim0.002$ could be the limit
of \bh evolution by Eddington-limited accretion enhanced due tomergers. 

\section{Summary}
We examine the discrepancy between the \bh - bulge luminosity relation of 
MBHs in normal galaxies (found by kinematic methods) and \rev -mapped 
Seyfert nuclei (Wandel 1999) for
updated data and find that the difference is smaller than first estimated 
but still significant.
We point out that although for the correlation between 
\bh mass and bulge velocity dispersion Seyferts and kinematicly detected
MBHs seem to agree, the AGN case in this correlation needs further study.
We suggest three classes of explanations for the apparent discrapency between the 
\bh mass - bulge luminosity relation of Seyfert galaxies and MBHs in normal galaxies:
Observational or method-related errors or bias,
Orientation-related effets for Seyferts 
and Evolution.
In reality, several of these causes may be present and contribute to the difference
between the \bh mass - bulge luminosity relations of 
Seyferts, quasars and MBHs in non-active galaxies.


\end{document}